\documentclass[prc,aps,float,nofootinbib,superscriptaddress,twocolumn]{revtex4-1}
\usepackage{amssymb}
\usepackage{amsmath}
\usepackage{amsfonts}
\usepackage{epsfig}
\usepackage{color} 
\usepackage{bbm}
\usepackage{comment}

\usepackage{multirow}
\usepackage{longtable}
\usepackage{array}
\usepackage{subfigure}
\usepackage{graphics}

\newcommand{\hcal}{{\cal H}}
\newcommand{\ncal}{{\cal N}}
\newcommand{\ocal}{{\cal O}}
\newcommand{\pcal}{{\cal P}}
\newcommand{\rcal}{{\cal R}}

\begin{document}

\title{Combining symmetry breaking and restoration with configuration interaction: \\ extension to z-signature symmetry in the case of the Lipkin Model}
  
\author{J. Ripoche} \email{julien.ripoche@cea.fr}
\affiliation{CEA, DAM, DIF, F-91297 Arpajon, France}

\author{T. Duguet} \email{thomas.duguet@cea.fr} 
\affiliation{KU Leuven, Instituut voor Kern- en Stralingsfysica, 3001 Leuven, Belgium}
\affiliation{CEA, DRF/IRFU/SPhN, Universit\'e Paris-Saclay, 91191 Gif-sur-Yvette, France}
\affiliation{National Superconducting Cyclotron Laboratory and Department of Physics and Astronomy, Michigan State University, East Lansing, MI 48824, USA}

\author{J.-P. Ebran} \email{jean-paul.ebran@cea.fr}
\affiliation{CEA, DAM, DIF, F-91297 Arpajon, France}

\author{D. Lacroix} \email{lacroix@ipno.in2p3.fr}
\affiliation{Institut de Physique Nucl\'eaire, IN2P3-CNRS, Universit\'e Paris-Sud, Universit\'e Paris-Saclay, F-91406 Orsay Cedex, France}

\date{\today}

\begin{abstract} 
\begin{description}
\item[Background] \textit{Ab initio} many-body methods whose numerical cost scales polynomially with the number of particles have been developed over the past fifteen years to tackle closed-shell mid-mass nuclei. Open-shell nuclei have been further addressed by implementing variants based on the concept of spontaneous symmetry breaking (and restoration). These methods typically access ground states properties and some restricted aspects of spectroscopy.
\item[Purpose]  In order to access the spectroscopy of open-shell nuclei more systematically while controlling the numerical cost, we design a novel many-body method that combines the merit of breaking and restoring symmetries with those brought about by low-rank individual excitations. 
\item[Methods] The recently proposed truncated configuration-interaction method based on optimized symmetry-broken and -restored states is extended to the $SU(2)$ group associated with total angular momentum. Dealing more specifically with the Lipkin Hamiltonian, the present study focuses on the breaking and the restoration of the z-signature symmetry associated with a discrete subgroup of $SU(2)$. The highly-truncated $N$-body Hilbert subspace within which the Hamiltonian is diagonalized is spanned by a z-signature broken and restored Slater determinant vacuum and associated low-rank, e.g. one-particle/one-hole and two-particle/two-hole, excitations. Furthermore, the extent by which the symmetry-unrestricted vacuum breaks z-signature symmetry is optimized {\it in presence} of projected low-rank particle-hole excitations. The quality of the method is gauged against exact ground- and excited-state eigenenergies for a large range of 
values of the two-body interaction strength. Furthermore, results are compared to those obtained from the generator coordinate method, the random-phase approximation and the self-consistent (second) random-phase approximation.
\item[Results] The proposed method provides an excellent reproduction of the ground-state energy and of low-lying excitation energies of various z-signatures and total angular momenta across the full range of inter-nucleon coupling defining the Lipkin Hamiltonian and driving the normal-to-deformed quantum phase transition. In doing so, the successive benefits of (i) breaking the symmetry, (ii) restoring the symmetry, (iii) including low-rank particle-hole excitations and (iv) optimizing the amount by which the underlying vacuum breaks the symmetry are illustrated. While the generator coordinate method building on the same deformed vacua provides results of similar quality, it is not the case of the symmetry-restricted random-phase and self-consistent (second) random-phase approximations in the strongly-interacting regime.
\item[Conclusions] The numerical cost of the newly designed variational method is polynomial with respect to the system size. It achieves a good accuracy on the ground-state energy and the low-lying spectroscopy for both weakly- and strongly-interacting systems. The present study confirms the results obtained previously for the attractive pairing Hamiltonian in connection with the breaking and restoration of $U(1)$ global gauge symmetry. These two studies constitute a strong motivation to apply this method to realistic nuclear Hamiltonians in view of providing a complementary, accurate and versatile \textit{ab initio} description of mid-mass open-shell nuclei. 
\end{description}
\end{abstract}



\maketitle

\section{Introduction}
\label{Intro}

The capacity of \textit{ab initio} many-body methods to describe more and more nuclei has grown tremendously over the last fifteen years~\cite{navratil00a,pieper02a,kowalski04a,hagen14a,roth07a,epelbaum10a,dickhoff04a,cipollone13a,hergert16a,hergert17a,launey16a,gebrerufael16a,tichai17a}. This results from the combined benefit of novel formal developments and increased computational capabilities. In particular, open-shell systems have recently been addressed by relying on the concept of spontaneous breaking (and possible restoration) of symmetries~\cite{soma11a,soma14b,henderson14a,duguet15a,signoracci15a,duguet17a}. Resulting methods, which scale polynomial with the number of particles, typically tackle ground-state properties at first while relying on further extensions/steps to access spectroscopy. A different strategy consists of targeting a closed-shell system to produce effective operators at play in a second-step valence-space configuration-interaction calculation, thus accessing open-shell systems' spectroscopy~\cite{bogner14a,jansen14a,dikmen15a}. While very powerful and successful, this method inherently carries the factorial scaling of the valence-space diagonalization. 

To access spectroscopy of (yet heavier) open-shell nuclei in a systematic and numerically controlled fashion, the present work wishes to combine the merit of breaking and restoring symmetries with features brought about by low-rank individual excitations. The highly-truncated $N$-body Hilbert subspace within which the Hamiltonian is diagonalized is spanned by a symmetry-broken and -restored vacuum along with associated low-rank individual excitations. Furthermore, the extent by which the underlying vacuum breaks the symmetry is optimized {\it in presence} of symmetry projected low-rank excitations. The fact that the underlying vacuum does break the symmetry in an optimal fashion allows the drastic limitation to low-rank individual excitations, thus strongly tempering the factorial scaling. The above elements bear strong resemblance with those that underlined the development of the variation after mean-field projection in realistic model spaces (VAMPIR) method~\cite{schmid84a} and of the projected shell model~\cite{hara95a} about thirty years ago. While there are several tiny differences, the main one relates to the fact that the presently proposed method is designed to apply to the full single-particle Hilbert space in conjunction with realistic inter-nucleon interactions. The approach laid out above also entertains parentage but significant differences with those designed in Refs.~\cite{allaart88a,caprio12a}.

Building the strategy initiated in Refs.~\cite{lacroix12a,gambacurta12a}, the above scheme was applied to tackle the exactly solvable attractive pairing Hamiltonian in Ref.~\cite{ripoche17a}. The symmetry of interest in this case is $U(1)$ global gauge symmetry associated to particle-number conservation. Results were very encouraging given the high-accuracy reproduction of both ground- and excited eigenenergies. Following the same philosophy, De Baerdemacker and collaborators have recently developped a method called Richardson-Gaudin Configuration-Interaction in Ref.~\cite{baerdemacker17a}. In order to validate the method with respect to yet another symmetry, the goal of the present paper is to apply it to the exactly solvable Lipkin model~\cite{lipkin65a} via the breaking and the restoration of the z-signature symmetry associated with a discrete subgroup of $SU(2)$. The objective is to compare the results to exact ones as well as to those obtained from the generator coordinate method, the random-phase approximation and the self-consistent (second) random-phase approximation.

This paper is organized as follows: Section~\ref{sec:formalism} displays the formalism in such a way that several standard methods can be recovered as particular cases. Sections~\ref{sec:potentialenergy}-\ref{sec:gcmresults} provide numerical results and compare them with exact solutions as well as results obtained from other existing approximate methods. Section~\ref{sec:conclusions} concludes the present work.

\section{Formalism}
\label{sec:formalism}

\subsection{Lipkin model}
\label{sec:lipkin}

The Lipkin model is an exactly solvable model often used to benchmark approximations of the nuclear many-body problem~\cite{meshkov65a,glick65a,agassi66a,schuck73a,holzwarth73a,ring80a,zhang90a,dukelsky90a,dukelsky91a,robledo92a,dukelsky99a,delion05a,severyukhin06a,hermes17a,wahlen17a}. This model features $N$ fermions distributed onto two $N$-fold degenerated shells separated by an energy $\varepsilon$. Each single-particle state is characterized by two quantum numbers $(p,\sigma)$. Within a shell, the value of $p$ differentiates the $N$ states. When belonging to the upper (lower) shell, each such state labelled by $p$ comes with $\sigma=+$ ($\sigma=-$). A two-body interaction is assumed to scatter pairs of particles from one shell to the other without changing the corresponding $(p,p')$ values. The associated Hamiltonian writes in second-quantized form as
\begin{equation}
    \label{lipkinhamiltonian}
    H \equiv \frac{\varepsilon}{2} \sum_{p\sigma} \sigma a^\dagger_{p\sigma}a_{p\sigma} - \frac{V}{2} \sum_{pp'\sigma}a^\dagger_{p\sigma}a^\dagger_{p'\sigma}a_{p'-\sigma}a_{p-\sigma},
\end{equation}
where $\{a^\dagger_{p\sigma}\}$ denotes creation operators associated with one-body basis states whereas  $V$ characterizes the interaction strength.

The unperturbed ground state of the system ($V=0$) displays $N$ particles in the lower shell, each particle occupying a state with a different value of the quantum number $p$, i.e. it is given by the Slater determinant
\begin{equation}
    \label{eq:hfsd}
    |\Phi\rangle \equiv \prod_{p=1}^N a^\dagger_{p-}|0\rangle \, .
\end{equation}
The interaction mixes $|\Phi\rangle$ with excited Slater determinants in which single-particle states characterized by the $N$ different $p$ values are always all occupied but in such a way that they can arbitrarily belong to the upper or the lower shells. There is a total of $2^N$ such many-body states, which makes solving the corresponding eigenvalue-problem of exponential cost when $N$ grows. Thankfully, symmetries of the Hamiltonian tremendously reduce the dimensionality one must effectively deal with.

\subsection{Quasi-spin algebra}
\label{sec:quasispin}

The resolution of the Lipkin model is based on the observation that bilinear products of creation and annihilation operators can be viewed as generators of Lie groups. In the present case, the Hamiltonian can be expressed in terms of quasi-spin operators and exact solutions can be obtained in a reduced many-body space. The components of the total quasi-spin operators of the system are defined as
\begin{subequations}
\label{eq:j0pmexpression}
\begin{align}
\label{eq:j0expr}
J_0 &\equiv \frac{1}{2} \sum_{p\sigma} \sigma a^\dagger_{p\sigma}a_{p\sigma} = J_z,  \\
J_+ &\equiv \sum_{p}a^\dagger_{p+}a_{p-} \, , \\
\label{eq:jmexpr}
J_- &\equiv \sum_{p}a^\dagger_{p-}a_{p+} \, ,
\end{align}
\end{subequations}
and obey usual commutation relations of the angular-momentum Lie algebra. 
Eventually, the Hamiltonian can be expressed as
\begin{equation}
    \label{eq:lipkinhamiltonianquasispin}
    \frac{H}{\varepsilon} = J_z - \frac{\chi}{2(N-1)} (J_+^2 + J_-^2) \, ,
\end{equation}
where the reduced interaction strength $\chi\equiv V(N-1)/\varepsilon$ has been introduced. In the following, we set $\varepsilon$ to $1$, which is equivalent to evaluating energies in units of $\varepsilon$.

The Lipkin Hamiltonian commutes with $J^2$ but not with $J_z$. Rather, $H$ commutes with the z-signature operator
\begin{equation}
    \label{eq:zsignatureoperator}
    R_z \equiv e^{i \phi}e^{i\pi J_z} \, ,
\end{equation}
which is defined up to a phase $\phi$. In order for the z-signature eigenvalue of the $N$-body ground-state of interest to be  $\eta = +1$, the phase is chosen as $\phi \equiv \pi N/2$. The other eigenvalues of $R_z$ are $\eta = -1$ and $\eta = \pm i$.

\subsection{Exact eigenstates}
\label{sec:exact}

Let $|\Psi^{JM}\rangle$ be eigenstates of $J^2$ and $J_z$. By construction, they are also eigenstates of $R_z$ with eigenvalue $\eta_M = (-1)^{N/2+M}$. Exact eigenstates $|\Xi^{J\eta}_k\rangle$ of $H$ being  labelled by eigenvalues $J$ and $\eta$, respectively associated with $J^2$ and $R_z$, can be represented as a linear combination of $|\Psi^{JM}\rangle$ states with the same $J$ value and $M$ values differing by an even number such that  $\eta_M=\eta$, i.e.
\begin{equation}
\label{eq:exacteigenstates}
|\Xi^{J\eta}_k\rangle \equiv \sum_{M|\eta_M=\eta} c^{JM\eta}_k |\Psi^{JM}\rangle
\end{equation}
where $k$ labels the eigenstates, for a given $J$ and $\eta$, from the lowest ($k=0$) to the highest ($k=J$ or $J-1$) 
energy. Ultimately, for integer values of $J$, the exact eigenstates with $\eta=+$ (resp. -) is obtained by 
diagonalizing the Hamiltonian in a space of size $(J+1)$ (resp. $J$).  

\subsection{Many-body basis}

\begin{table}
  \setlength\extrarowheight{2pt}
  \begin{tabular}{c|l}
    \hline
    \hline
    GS & Ground State \\
    \hline
    IRREP & Irreducible Representation \\
    \hline
    RHF & Restricted Hartree-Fock \\
    \hline
    UHF & Unrestricted Hartree-Fock \\
    \hline
    PAV & Projection After Variation \\
    \hline
    PHF & Projected Hartree-Fock \\
    \hline
    VAP & Variation After Projection \\
    \hline
    RVAP & Restricted Variation After Projection \\
    \hline
    RPA & Random-Phase Approximation \\
    \hline
    SCRPA & Self-Consistent Random-Phase Approximation \\
    \hline
    TCI & Truncated Configuration Interaction \\
    \hline
    GCM & Generator Coordinate Method \\
    \hline
    GTCI & Generalized Truncated Configuration Interaction \\
    \hline
    \hline
  \end{tabular}
  \caption{\label{acronymtable} Table of acronyms used in the text.}
\end{table}

The present many-body method is based on an ansatz that expands approximate eigenstates on a set of states that we now introduce.  The goal is to explore, starting from symmetry-restricted or symmetry-unrestricted Hartree-Fock (HF) theory, different techniques to construct a highly-truncated set of many-body states used to diagonalize the Hamiltonian. 

\subsubsection{Symmetry-restricted basis states}

Let us first consider the restricted HF (RHF) approximation to the $N$-body ground-state. It is nothing but the Slater determinant $| \Phi \rangle$ introduced in Eq.~\ref{eq:hfsd} with an associated RHF energy equal to $-N/2$. It can be checked that the RHF state is a specific eigenstate of $(J^2,J_z)$ characterized by $| \Phi \rangle = | \Psi^{J-J} \rangle$ with $J=N/2$.

From the RHF vacuum $| \Phi \rangle$, n-particle/n-hole (npnh) excitations are generated according to
\begin{equation}
\label{npnh}
| \Phi^{p_1\ldots p_n} \rangle = \prod_{k=1}^n a^\dagger_{p_k+}a_{p_k-}| \Phi \rangle \, .
\end{equation}
These excited Slater determinants are not eigenstates of  $(J^2,J_z)$ and it is, thus, more convenient to consider specific linear combinations of them that do have such property, i.e. the linear combinations of npnh excitations ($n \leq 2J$) of the type
\begin{align}
    \label{eq:hfjmcorrespondance}
    |\Psi^{J-J+n}\rangle &= J_{+}^{n} | \Phi \rangle
    = \frac{1}{\sqrt{\binom{N}{n}}} \sum_{p_1 < ... < p_n}| \Phi^{p_1\ldots p_n} \rangle \, .
\end{align}

\subsubsection{Symmetry-breaking basis states}

While the above set of states is built from the RHF Slater determinant, a more general starting point consists of allowing the reference vacuum to break  z-signature symmetry~\cite{agassi66a}, i.e. to allow the solution of the HF equations to have a (chosen) deformation with respect to the corresponding subgroup of SU(2). The symmetry unrestricted Hartree-Fock (UHF) Slater determinant reads as
\begin{equation}
    \label{eq:defhf}
    |\Phi(\Omega)\rangle \equiv R(\Omega) |\Phi\rangle = \prod_{p=1}^N a^\dagger_{p-}(\Omega)|0\rangle \, ,
\end{equation}
where creation and annihilation operators have been rotated in quasi-spin space according to
\begin{subequations}
\label{eq:quasispinrotation}
\begin{align}
        a^\dagger_{p\sigma}(\Omega) &\equiv R^\dagger(\Omega) a^\dagger_{p\sigma} R(\Omega) \, , \\
        &= \cos(\alpha) a^\dagger_{p\sigma} -\sigma \sin(\alpha)e^{-i \sigma \varphi} a^\dagger_{p-\sigma} \, .
\end{align}
\end{subequations}
In the above equation, the rotation operator $R(\Omega)$ reads
\begin{equation}
    R(\Omega) \equiv \exp \left(- \Omega J_+ + \Omega^* J_- \right) \, ,
\end{equation}
where the variable $\Omega \equiv \alpha e^{i\varphi}$ actually collects two rotation ("deformation") parameters.

Starting from the set of states $|\Psi^{JM}\rangle$ introduced in Eq.~\eqref{eq:hfjmcorrespondance}, the application of $R(\Omega)$ delivers a new set of configurations
\begin{equation}
    \label{eq:uhfsd}
    |\Psi^{JM}(\Omega)\rangle \equiv R(\Omega) |\Psi^{JM}\rangle \, ,
\end{equation}
corresponding to linear combinations of npnh excitations ($M>-N/2$) on top of the UHF Slater determinant ($M = -N/2$). While states $|\Psi^{JM}(\Omega)\rangle$ carry labels $(J,M)$ as a memory of the states they originate from, they are neither eigenstates of $J_z$, nor of $R_z$.

\subsubsection{Symmetry-breaking and -restored basis states}

Since exact eigenstates do carry good z-signature $\eta$, a more appropriate set of states to expand our approximate eigenstates are obtained by further applying the projection operator
\begin{equation}
    \label{eq:projector}
    P_{\eta} \equiv \frac{1}{2} \left( 1 + \eta R_z \right) \, ,
\end{equation}
thus, reading
\begin{align}
    \label{eq:symresstates}
    |\Theta^{JM\eta} (\Omega) \rangle &\equiv P_{\eta} |\Psi^{JM} (\Omega) \rangle \nonumber \\
    &= \frac{1}{2} \left( |\Psi^{JM} (\Omega) \rangle + \eta \eta_M |\Psi^{JM} (-\Omega) \rangle \right) \, .
\end{align}
For a given value of the deformation $\Omega$, $\{|\Theta^{JM\eta} (\Omega) \rangle, M = -J,\ldots,+J\}$ constitutes a set ${\cal B}^{J\eta}(\Omega)$ of non-orthogonal states with good angular momentum $J$ and z-signature $\eta$ that results from projecting deformed (linear combinations of) npnh excitations on top of the UHF Slater determinant $|\Phi(\Omega)\rangle$. The quantum number $M$ labels the state $|\Psi^{JM} \rangle$ the final state originates from but does {\it not} mean that $|\Theta^{JM\eta} (\Omega) \rangle$ is an eigenstate of $J_z$. As a matter of fact, a z-signature $\eta=\pm 1$ can be obtained from an original state whose associated $M$ value corresponds to the opposite z-signature $\eta_M=\mp 1$ as soon as $\Omega \neq 0$. For $\Omega=0$, one recovers $|\Phi(0)\rangle = |\Phi\rangle$ and $|\Psi^{JM} (0) \rangle=|\Psi^{JM} \rangle$ such that the application of $P_{\eta}$ becomes superfluous.

\subsection{Generalized truncated CI method}
\label{sec:truncCI}

We wish to approximate eigenstates of $H$, starting with its ground state, via an exact diagonalization within a highly truncated subspace. In its most general setting, the subspace in question is spanned by the set of states $\{{\cal B}^{J\eta}(\Omega), |\Omega|\leq\pi/4\}$, i.e., exact eigenstates are approximated by the ansatz
\begin{equation}
    \label{eq:fullansatz}
    |\Lambda^{J\eta}_k\rangle \equiv \sum_{\Omega \in {\rm mesh}} \sum_{M=-J}^{M_{\text{max}}} c^{JM\eta}_k (\Omega) |\Theta^{JM\eta} (\Omega) \rangle \, ,
\end{equation}
where $M_{\text{max}} \leq J$ and where $\Omega \in {\rm mesh}$ designates the selected set of deformations included in the expansion. The ansatz introduced in Eq.~\eqref{eq:fullansatz} defines the so-called generalized truncated configuration interaction (GTCI) method. It constitutes a generalization of the TCI method applied to the pairing Hamiltonian in Ref.~\cite{ripoche17a} given that ansatz~\eqref{eq:fullansatz} allows the mixing of states associated with different deformations $\Omega$. 

The energy associated with state $|\Lambda^{J\eta}_k\rangle$ is given by\footnote{It is worth noting that the eigenstates and associated eigenenergies are function of the reduced interaction strength $\chi$ defining the Lipkin Hamiltonian $H$.}
\begin{equation}
  E^{J\eta}_k = \frac{\langle\Lambda^{J\eta}_k|H|\Lambda^{J\eta}_k\rangle}{\langle\Lambda^{J\eta}_k|\Lambda^{J\eta}_k\rangle} \, .
\end{equation}
Applying a variational principle to the trial state, coefficients of the mixing are obtained by solving a Schroedinger equation represented in a restricted non-orthogonal basis by
\begin{align}
    \label{eq:diago1}
    \sum_{\Omega' M'} c^{JM'\eta}_{k} (\Omega') \left( {\cal H}^{J\eta}_{MM'}(\Omega;\Omega') -   E^{J\eta}_k {\cal N}^{J\eta}_{MM'}(\Omega;\Omega') \right) = 0\, ,
\end{align}
where the expressions of the Hamiltonian and norm matrices, defined in each irreducible representation $(J,\eta)$, are provided in appendix. The eigenvalue problem is solved by first diagonalizing the norm matrix through a unitary transformation leading to a new set of orthonormal states that is eventually used to diagonalize $H$.

\subsection{Particular cases}
\label{subcases}

The above many-body scheme incorporates several existing approaches as particular cases
\begin{enumerate}
\item Considering a particular value for the deformation $\Omega_{\text{aux}}$ and setting $c^{JM\eta}_k (\Omega)$ to zero for $\Omega \neq \Omega_{\text{aux}}$ lead to using the reduced ansatz
\begin{equation}
  |\Lambda^{J\eta}_{k}\rangle  \equiv \sum_{M=-J}^{M_{\text{max}}} c^{JM\eta}_{k} (\Omega_{\text{aux}}) |\Theta^{JM\eta} (\Omega_{\text{aux}}) \rangle \, . \label{ansatz1}
\end{equation}
The configuration mixing is, thus, limited to npnh configurations without any fluctuation of the collective variable (i.e. deformation) $\Omega$. Such an ansatz denotes a purely {\it vertical} expansion. While ansatz~\eqref{ansatz1} provides exact solutions for every $\Omega_{\text{aux}}$ if $M_{\text{max}}=J$, our present goal is to employ it for $M_{\text{max}}\ll J$. In this case, and as numerically exemplified later on, the quality of the purely vertical expansion strongly depends on the  value $\Omega_{\text{aux}}$ of the collective deformation. Naively, $\Omega_{\text{aux}}$ can be set to zero, in which case Eq.~\eqref{ansatz1} reduces to a standard, symmetry restricted, (truncated) CI approach. Rather, $\Omega_{\text{aux}}\equiv\Omega_{\text{UHF}}$ is defined at the minimum of the UHF energy while $\Omega_{\text{aux}}\equiv\Omega_{\text{RVAP}}$ is obtained at the minimum after the symmetry restoration has been applied (i.e. the so-called restricted variation after projection (RVAP) minimum~\cite{rodriguez05a}). Eventually, $\Omega_{\text{aux}}\equiv\Omega_{\text{opt}}$ corresponds to a full optimization of the deformation in presence of both the symmetry restoration {\it and} the npnh excitations included in the variational ansatz. The latter option corresponds to the choice made in Ref.~\cite{ripoche17a} and defines the so-called TCI method.
\item Setting $c^{JM\eta}_k (\Omega)$ to zero for $M \neq -J$ leads to using the reduced ansatz
\begin{equation}
  |\Lambda^{J\eta}_{k}\rangle \equiv \sum_{\Omega \in {\rm mesh}} c^{J-J\eta}_{k} (\Omega) |\Theta^{J-J\eta} (\Omega) \rangle \, , \label{ansatz2}
\end{equation}
corresponding to the adiabatic symmetry-restored generator coordinate method (GCM). The configuration mixing is thus limited to fluctuations of the collective deformation $\Omega$ defining the (constrained) UHF vacuum and does not incorporate any individual npnh excitation. Such an ansatz denotes a purely {\it horizontal} expansion. For the Lipkin Hamiltonian, the GCM ansatz provides exact solutions using an equidistant spacing mesh when the number of states matches the full dimensionality of the problem~\cite{ring80a}. It is also worth noting that the GCM ansatz itself can restore the z-signature symmetry such that the explicit projection becomes superfluous. Indeed, an eigenstate of $R_z$ with eigenvalue $\eta=+$ (resp. $\eta=-$) is obtained by mixing symmetry breaking states $|\Psi^{J-J} (\pm\Omega) \rangle$ in ansatz~\eqref{ansatz2} with equal (resp. opposite) weights $c^{J-J}_{k} (-\Omega) = c^{J-J}_{k} (\Omega)$ (resp. $c^{J-J}_{k} (-\Omega) = -c^{J-J}_{k} (\Omega)$).
\end{enumerate}

The rationale behind the full ansatz introduced in Eq.~\eqref{eq:fullansatz} is to eventually achieve the best possible optimization of the highly-truncated subspace employed by combining the benefits of both horizontal and vertical expansions. In the present case, however, the Lipkin Hamiltonian happens to be too simplistic to allow us to test such a highly-tuned ansatz. We, thus, presently limit ourselves to testing the purely horizontal and the purely vertical expansions separately. It already enables the characterization of the merits of each of these two expansions in reproducing exact solutions at the lowest possible cost. When dealing with a richer Hamiltonian later on, the optimized combination of both expansions can be envisioned and tested.

\section{Potential energy}
\label{sec:potentialenergy}

\begin{figure}[h!]
    \centering
    \includegraphics[scale=1.0]{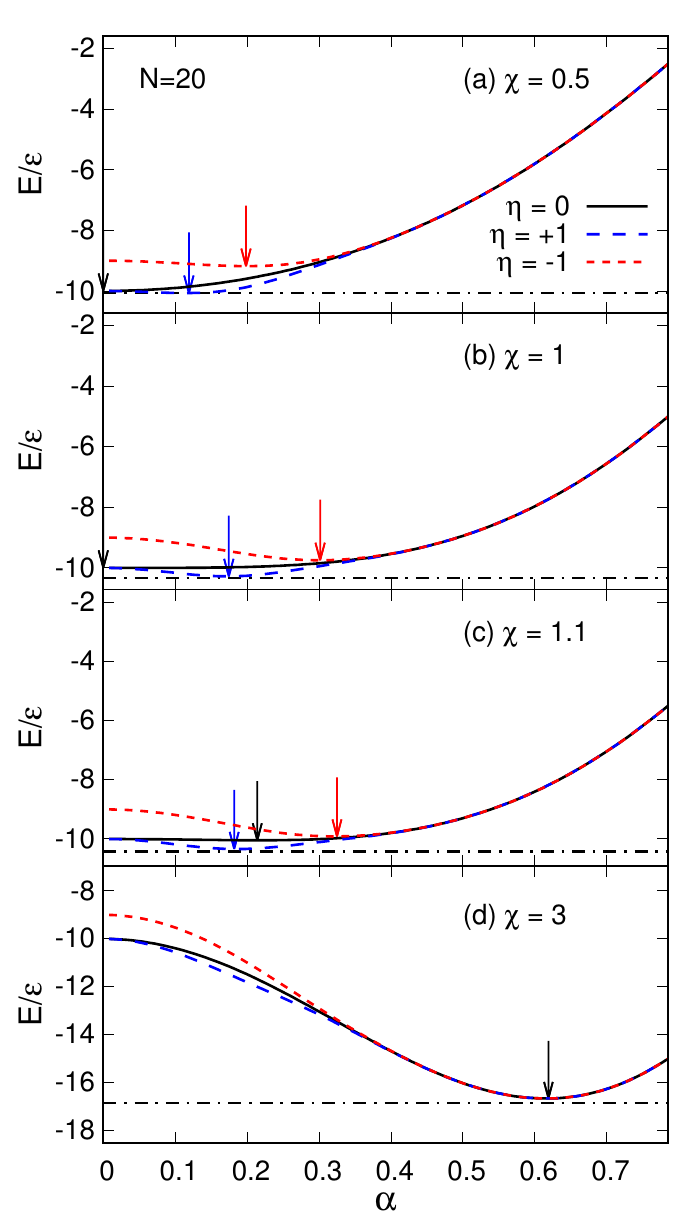}
    \caption{(Color online) Potential energy curves (in units of $\varepsilon$) as a function of $\alpha$ for $N=20$ and  (a) $\chi=0.5<\chi_c$, (b) $\chi=1\equiv\chi_c$, (c) $\chi=1.1>\chi_c$ and (d) $\chi=3.0\gg\chi_c$. The UHF (black solid line) along with the positive (blue dashed line) and negative (red short dashed 
  line) z-signature projected energies are displayed in each case. Arrows point to the global minimum of each curve. A single arrow appears in panel (d) given that the global minimum is the same for the three curves in this case. In all panels the horizontal black dotted-dashed line corresponds to the exact ground-state energy. }
    \label{fig:figure7}
\end{figure}

As a precursor of TCI and GCM calculations, we presently discuss projected Hartree-Fock (PHF) potential energy curves as a function of the deformation $\Omega$
\begin{subequations}
\label{eq:potentialenergy}
\begin{align}
    E^{\text{PHF}}_{\eta}(\Omega) &\equiv \frac{\langle \Theta^{J-J\eta} (\Omega)| H | \Theta^{J-J\eta} (\Omega) \rangle }{\langle \Theta^{J-J\eta} (\Omega)| \Theta^{J-J\eta} (\Omega) \rangle} \\
    &= \frac{\langle \Phi (\Omega)| H P_{\eta} | \Phi (\Omega) \rangle }{\langle \Phi  (\Omega)| P_{\eta} | \Phi  (\Omega) \rangle} \, ,
\end{align}
\end{subequations}
for $\eta = \pm$. Further setting $\eta = 0$ sends back the unprojected potential energy, i.e., the UHF potential energy curve from which the RHF energy is extracted at $\varphi=\alpha=0$. The potential energy curves are shown in Fig.~\ref{fig:figure7} as a function of $\alpha$ (for $\varphi=0$) for four values of the reduced interaction strength $\chi$. 

For $\chi<\chi_c=1.0$ (panel (a)), the minimum of the UHF curve is obtained for the RHF solution ($\varphi=\alpha=0$), i.e. the system is in a normal phase and no spontaneous symmetry breaking occurs at the mean-field level. For $\chi>\chi_c$ (panels (c) and (d)), the minimum of the UHF curve is obtained for $\varphi=0$\footnote{As $\varphi$ does not play a crucial role, it is set to zero in the present work. See Ref.~\cite{severyukhin06a} for a corresponding discussion.} and $\alpha=\arccos(1/\chi)/2$ (e.g. $\alpha=0.615$ in panel (d)), i.e. the system undergoes a spontaneous symmetry breaking at the mean-field level.  For $\chi=\chi_c$ (panel (b)), the system is unstable with respect to deformation and is exactly on the edge of undergoing the phase transition eventually observed in panels (c) and (d). In all cases, the RHF energy ($\alpha=0$) is equal to $-N/2=-10$. While in panel (c) the symmetry breaking is weak and brings little energy gain as compared to the RHF reference, the UHF minimum is significantly lower than it in panel (d) where the symmetry breaking is strong such that a large amount of static correlations is captured to approach the exact ground-state energy.

The PHF state with positive z-signature approximates the ground state, i.e., the lowest state in the irreducible representation (IRREP) characterized by $J=N/2=10$ and $\eta=+1$. The corresponding energy curve is systematically below or superimposed on the UHF one. At the UHF minimum, there is no gain in energy in panels (a), (b) and (d), i.e., the projection after variation (PAV) method does not provide additional correlation energy beyond UHF in these cases. In the first two cases, as explained above, the UHF minimum does not differ from the RHF one (i.e. it occurs at $\varphi=\alpha=0$) such that no symmetry is broken in the first place. In panel (d), on the contrary, the symmetry is so strongly broken at the UHF minimum (i.e. it occurs at a large $\alpha$ value) that the projection, in spite of restoring the good z-signature quantum number in the ground state, does not bring any additional correlation energy. Indeed, the two components appearing in Eq.~\eqref{eq:symresstates} are essentially orthogonal and not coupled by the Hamiltonian in Eq.~\eqref{eq:potentialenergy}. The situation is qualitatively different in panel (c) (actually on the whole interval $1<\chi<1.5$) where the symmetry breaking is weak at the UHF minimum. In this case, the configuration mixing associated with the symmetry projection does bring additional correlation energy, i.e. the PHF curve is below the UHF one at the UHF minimum. More interestingly, the RVAP energy corresponding to the global minimum of the PHF energy curve captures additional static correlations beyond UHF in panels (a), (b)  and (c), i.e., when the latter does not spontaneously break the symmetry or only breaks it weakly. In these cases, the PHF minimum differs from the UHF one. However, when the z-signature is strongly broken at the UHF level (panel (d)), static correlations are fully grasped at that level and the projection does not lower the energy any further as explained above. 

The PHF state  with negative z-signature approximates the lowest state in the IRREP characterized by $J=N/2=10$ and $\eta=-1$. The corresponding energy curve is systematically above or superimposed with the UHF one. In panels (a), (b) and (c), the global minimum is at higher energy and larger deformation than for the positive z-signature curve. In panel (d) where the symmetry is strongly broken, both global minima are the same. One can, thus, anticipate that the energy difference between the first positive and negative z-signature states will decrease as the strength increases. Correspondingly, the first negative z-signature excited state will be at lower energy in the deformed phase than in the normal phase.  

\section{Vertical expansion (TCI)}
\label{sec:ciresults}

We are now in position to perform TCI calculations based on a pure vertical expansion of the many-body states (see Eq.~\eqref{ansatz1}). In the following, the notation $(i+j+...)\text{ph}_{\alpha}$ is employed to specify that the configurations used in the vertical expansion are ipih, jpjh, ..., on top of the UHF vacuum with deformation parameter $\alpha$. For instance, $(0)\text{ph}_{\alpha}$ means that only the vacuum is included while $(0+1+2)\text{ph}_{\alpha}$ means that 1p1h and 2p2h states are further added. The subscript $\eta$ is added whenever the configurations have been projected onto good z-signature.

\subsection{Ground state}
\label{sec:ciresultsgs}

\begin{figure}[h!]
    \centering
    \includegraphics[scale=1.0]{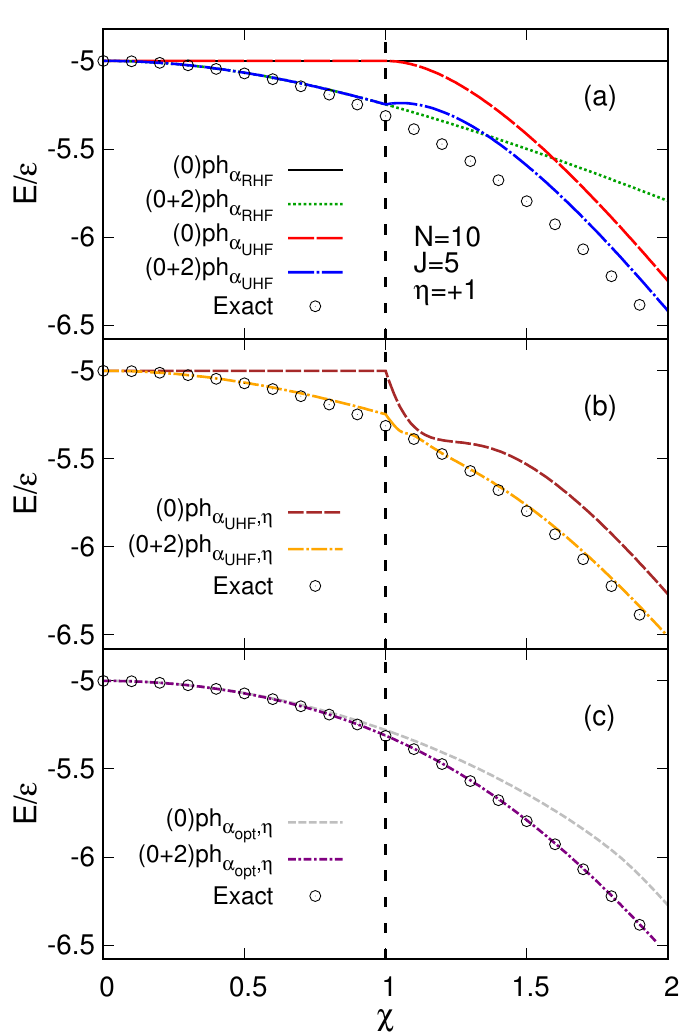}
  \caption{(Color online) Absolute ground-state energy as a function of the coupling strength $\chi \in [0,2]$ for $N=10$ and $J=5$. The three panels compare successive levels of approximations to exact results. Panel (a) RHF and UHF with and without inclusion of associated 2p2h configurations. Panel (b) PHF with and without inclusion of 2p2h configurations all at the deformation of the UHF minimum.
  Panel (c) TCI results obtained by optimizing the $\alpha$ parameter with and without inclusion of  2p2h configurations. 
  Note that, when low-rank excitations are included, the optimization is made in the presence of them.
  The critical value $\chi_c= 1$ is indicated by the black dashed vertical line.}
    \label{fig:figure1}
\end{figure}

Fig.~\ref{fig:figure1} illustrates the successive benefits of breaking the z-signature symmetry (panel (a)), of restoring it (panel (b)) and of optimizing the symmetry breaking of the underlying vacuum (panel (c)) for the description of the ground-state energy. In each case, the impact of including low-rank, e.g., 2p2h, individual excitations is displayed as well. Results are shown for $N=10$ for a good visualization but the analysis of the results would be the same for different, e.g., $N=20$, particle numbers.

As seen in panel (a), the RHF solution is exact in the absence of two-body interaction ($\chi=0$) and a decent approximation in the very small coupling regime ($\chi<0.2$). As the Hamiltonian only couples states with the same z-signature, 2p2h configurations on top of the RHF state are the first to amend the latter, i.e., 1p1h excitations do not contribute. Adding 2p2h configurations provides a good reproduction of the exact solution over the entire normal phase ($\chi<\chi_c$) but not in the deformed phase ($\chi>\chi_c$), eventually leading to the wrong asymptotic behaviour for $\chi\gg\chi_c$. Contrarily, the UHF solution has a good asymptotic behaviour and already leads to a better approximation of the ground-state energy than the RHF solution with 2p2h excitations for $\chi>1.6$. It is, thus, clear that breaking the z-signature qualitatively improves the description beyond $\chi_c$, especially at very large coupling strength, even in the absence of projection. As for the RHF in the normal phase, it is however necessary to add 2p2h excitations on top of the UHF minimum to reach a decent accuracy over the whole deformed phase. Still, the description is far from perfect, especially in the weakly deformed regime and at the very phase transition where an artificial kink appears. 

As seen in panel (b), and as discussed in Sec.~\ref{sec:potentialenergy}, the PAV does not lower the energy in the normal phase or for very deformed systems. On the other hand, it does improve very significantly the description for $\chi_c<\chi<1.5$ at the price of inducing an artificial kink at the phase transition. At the UHF minimum, the projection performed in presence of 2p2h configurations improves the description very significantly over the deformed phase. The reproduction of the exact ground-state energy is now quantitatively satisfactory over the entire range of coupling values, except just before the phase transition. Qualitatively, the phase transition is still artificially provided with a first-order character, which is definitely absent from exact results.

\begin{figure}[h!]
    \centering
    \includegraphics[scale=1.0]{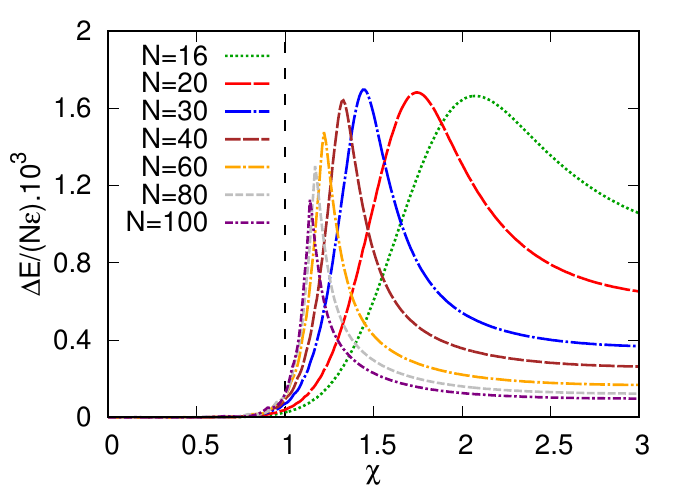}\\
    \includegraphics[scale=1.0]{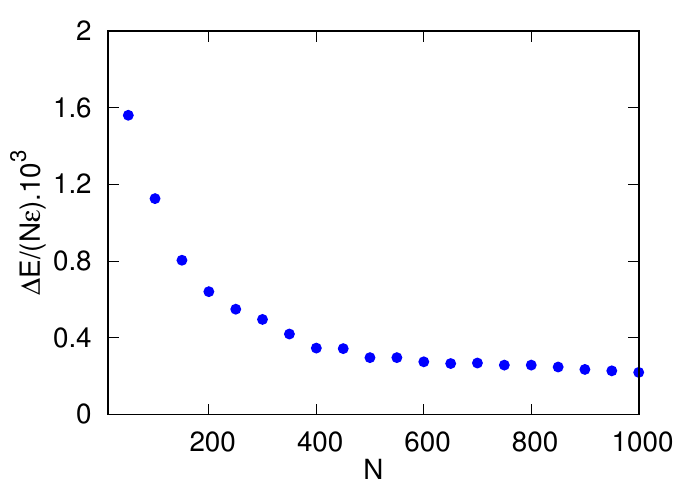}
  \caption{(Color online) Top: Error on the ground-state energy per particle  as a function of $\chi$ for $N=16,\dots,100$. 
  Results are obtained from the optimized TCI method including 1p1h and 2p2h configurations. Bottom: maximum error over the interval $\chi \in [0,3]$ on the ground-state energy per particle as a function of the particle number. }
    \label{fig:systematics}
\end{figure}

Panel (c) illustrates how optimizing the symmetry breaking of the underlying UHF vacuum in presence of the symmetry restoration and of 2p2h configurations\footnote{1p1h configurations are not included here as they only weakly couple to the PHF state (they entirely decouple when the symmetry is not restored). They will be included later on to generate fully optimized results.} does correct the remaining deficiencies. First of all, even when the UHF does not spontaneously break the symmetry ($\chi<\chi_c$), it does so when optimized in presence of the symmetry restoration. This leads to a tremendous improvement of the description over the normal phase, which eventually carries over to the deformed phase. This constitutes the strict RVAP result. Adding projected 2p2h configurations and further optimizing the deformation of the underlying vacuum eventually provides a nearly perfect reproduction of the exact ground-state energy throughout the entire range of coupling strengths. This constitutes an "advanced" RVAP result. As illustrated in Fig.~\ref{fig:systematics}, the absolute error on the ground-state energy per particle remains below $1.6 \, 10^{-3} \varepsilon$ (i.e. below $0.3\%$ relative error) for all particle numbers and over the full interval $\chi \in [0,3]$. In particular, the inadequate first-order character of the phase transition is fully amended. In addition, the reproduction of the ground-state energy improves as the particle number increases, i.e. as the size of the Hilbert space grows. The precision and scaling with particle number  is particularly remarkable in view of the small size of the highly-truncated many-body Hilbert space used here. This directly stems from the use of a strongly-entangled state optimized to incorporate the physics close to a quantum phase transition. The merits of the TCI method presently illustrated for the Lipkin Hamiltonian and the z-signature symmetry are in line with the results obtained for the pairing Hamiltonian and $U(1)$ global gauge symmetry~\cite{ripoche17a}.

\subsection{Excited states}
\label{sec:ciresultses}

Because the TCI method diagonalizes the Hamiltonian in a highly-optimized and truncated (non-orthogonal) basis, it naturally accesses low-lying excited states. Excitation energies are defined as
\begin{equation}
  E_{\text{exc},k}^{J \eta} = E_{k}^{J \eta} - E_{0}^{N/2 +}
\end{equation}
where $E_{0}^{N/2 +}$ and $E_{k}^{J \eta}$ denote the absolute ground-state energy ($\eta=+$) and the energy of the $k$-th excited state in the IRREP  ($J,\eta$), respectively.

\begin{figure}[h!]
    \centering
    \includegraphics[scale=1.0]{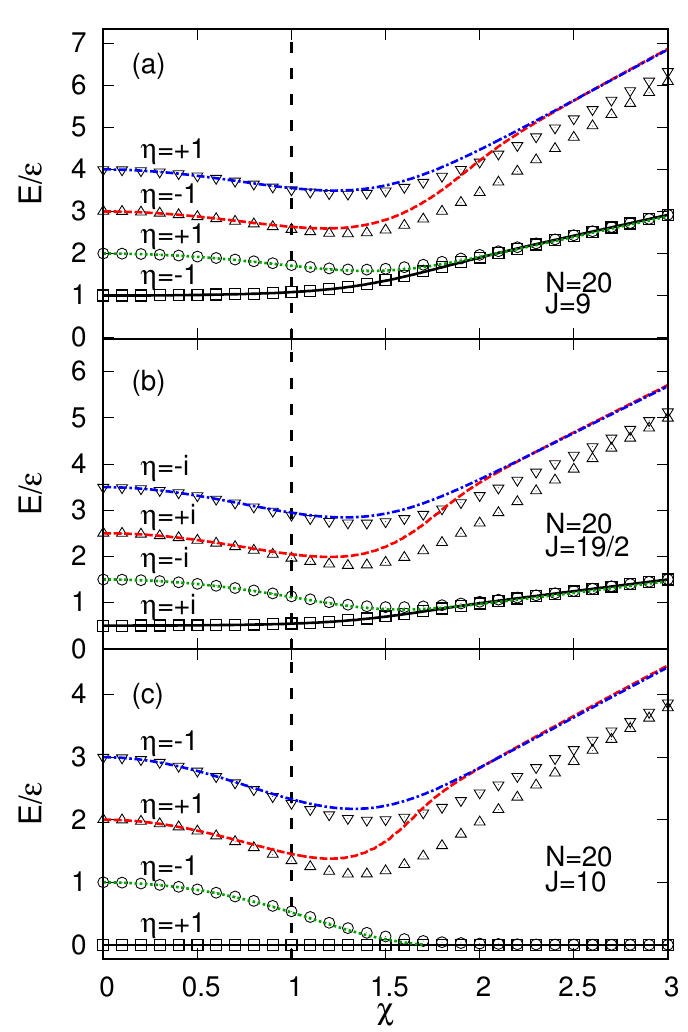}
  \caption{(Color online) Low-lying excitation energies for $N=20$ as a function of $\chi \in [0,3]$. Panel (a): $J=9$ and $\eta=\pm 1$. Panel (b): $J=19/2$ and $\eta=\pm i$. Panel (c): $J=10$ and $\eta=\pm 1$. Results from the TCI method including up to 2p2h excitations (lines) are compared to exact results (symbols). The deformation of the underlying vacuum is optimized  in each IRREP $(J,\eta)$ to minimize the energy of the lowest state.}
    \label{fig:figure8}
\end{figure}

Fig.~\ref{fig:figure8} compares the excitation energy of twelve low-lying states corresponding to three different $J$ values against exact results as a function of $\chi \in [0,3]$ for $N=20$. Within any given IRREP $(J,\eta)$, the deformation of the underlying vacuum is chosen to optimize the energy of the lowest state. Even if not based on the same vacuum, many-body states belonging to different IRREPs are orthogonal by virtue of their different symmetry quantum numbers.

\begin{figure}[h!]
    \centering
   \includegraphics[scale=1.0]{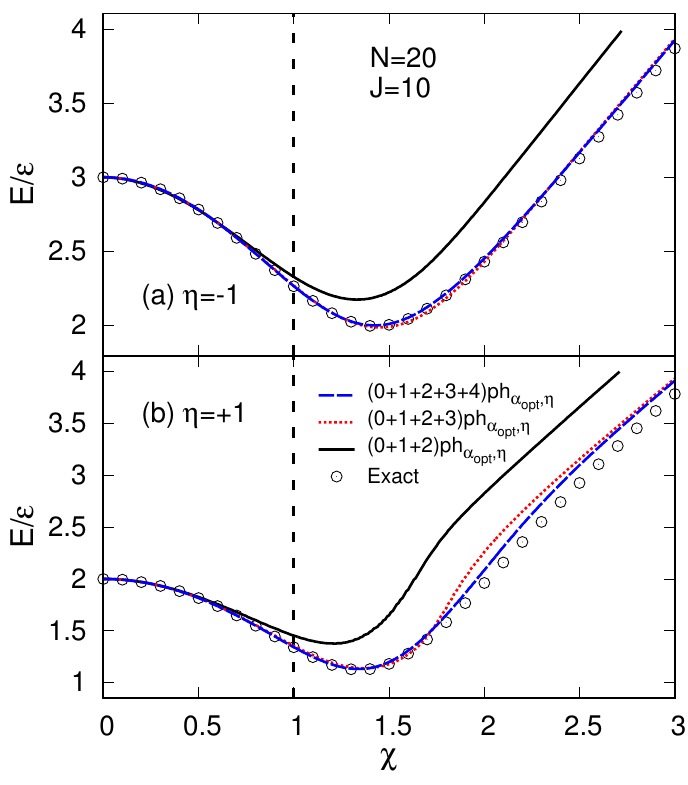}
  \caption{(Color online) Focus on the second (b) and third excited state (a) shown in panel (c) of Fig.~\ref{fig:figure8}. 
  The exact results (filled circles) are systematically compared to the optimized TCI method including
  up to  2p2h configurations (black solid line),  3p3h configurations (red dotted line) and
  4p4h configurations (blue dashed line).}
    \label{fig:3p3h}
\end{figure}

Panel (c) deals with IRREPs $(J=10,\eta=\pm)$ and includes the ground state whose energy is set to $0$ by definition. The first excited state has a negative z-signature. As anticipated in Sec.~\ref{sec:potentialenergy}, its excitation energy decreases steadily as a function of $\chi$ and becomes asymptotically degenerate with the ground state. The corresponding TCI result matches perfectly the exact energy for all values of the coupling strength, in part because the vacuum deformation was chosen to optimize the energy of that $(J=10$, $\eta=-)$ state. The two other $J=10$ states are fitly reproduced in the normal phase but quickly degrade in the deformed phase. They lie too high in energy and become degenerate at too small coupling strength. While static correlations are efficiently included via the breaking and the restoration of the z-signature symmetry, the reproduction of these excited states at large coupling is obtained via a better account of dynamical correlations. As demonstrated in Fig.~\ref{fig:3p3h}, the inclusion of 3p3h configurations allows one to fully capture the second negative z-signature state over the full range $\chi \in [0,3]$. While the second positive z-signature state is also very significantly improved, 4p4h configurations are necessary to postpone its degeneracy with the second negative z-signature state to the appropriate coupling value.

Similar observations hold true for panels (a) and (b). The energy of the lowest state is impeccably reproduced in each IRREP by virtue of optimizing the deformation of the underlying vacuum. Next states are also very well reproduced for all coupling strengths, except for the third and fourth excited states (those shown) that necessitate the inclusion of higher-rank npnh configurations at large couplings. All in all, the TCI method efficiently grasps static and dynamical correlations in both the normal and the deformed phases for both ground and low-lying states.

\begin{figure}[h!]
    \centering
    \includegraphics[scale=1.0]{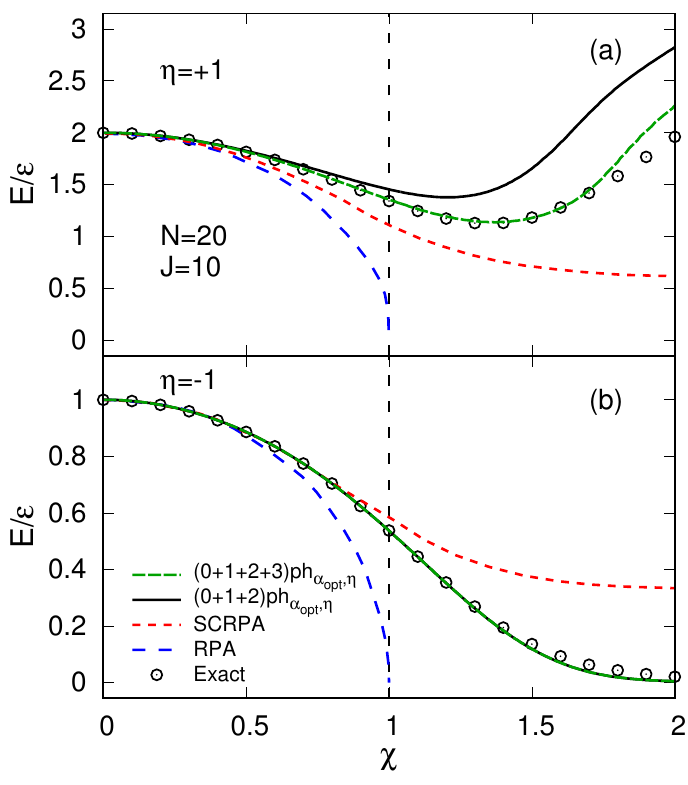}  
  \caption{(Color online) First excited state energy with $J=10$ and (a) $ \eta=+1$ or (b) $\eta=-1$ obtained for $N=20$ particles. Exact results (open circles) are compared to those obtained via RPA (blue dashed line) and SCRPA (red short dashed line) \cite{schuck73a} as well as via the TCI approach including up to 2p2h (black solid line) and 3p3h  (green dashed line) configurations.}
    \label{fig:excitationEnergiesRpa}
\end{figure}

In Fig.~\ref{fig:excitationEnergiesRpa}, the excitation energy of the first $(\eta=-)$ and second excited $(\eta=+)$ states with $J=10$ are compared for $N=20$ to results obtained from the random-phase approximation (RPA) and the self-consistent RPA (SCRPA)~\cite{schuck73a} performed on the RHF vacuum. The three methods are essentially exact in the weak coupling regime. This results from the fact that they all explicitly incorporate 1p1h and 2p2h configurations with respect to the RHF vacuum\footnote{At exactly zero interaction strength, the RPA and self-consistent RPA methods cannot address the second excited state that is a pure 2p2h excitation on top of the RHF vacuum. Contrarily, taking the limit $\chi\rightarrow 0$ does provide the correct value thanks to the fact that the RPA excitation operator acts on a correlated ground-state as soon as $\chi\neq 0$ and thus generates the 2p2h components necessary to describe the second excited state. There is thus a discontinuity at $\chi=0$ in the RPA calculation of that second excited state.} that dominate the structure of the lowest and second excited states, respectively, in the non-interacting limit. The excitation energy degrades as $\chi$ increases in the RPA and vanishes for $\chi=\chi_c$ such that there is no real solution for $\chi>\chi_c$. This relates to the known failure of the RPA with regards to dealing with phase transitions. To overcome this difficulty, a solution consists of enforcing  self-consistency via the SCRPA method. As shown in Fig.~\ref{fig:excitationEnergiesRpa}, the SCRPA is indeed able to reproduce the excitation energy all throughout the normal phase and to avoid  the collapse at $\chi=\chi_c$ and beyond. However, the description in the deformed phase ($\chi>\chi_c$) is qualitatively and quantitatively incorrect. In particular, the steady decrease of the excitation energy leading to a degeneracy with the ground state is not captured by the SCRPA based on the RHF Slater determinant. By comparison, the optimized symmetry breaking and restoration efficiently captures mandatory static correlations (along with dynamical correlations) in the TCI method beyond the phase transition. These correlations are hard to come by in RHF-based many-body methods. While it is significantly better than SCRPA beyond the phase transition, even the nonlinear higher RPA with truncation $d=2$ built on the RHF reference~\cite{terasaki17a}, which is equivalent to the self-consistent (extended) second RPA~\cite{schuck16a}, provides an accurate description of the first $\eta=-1$ state only up to $\chi \approx 1.14$ (see line B in panel (b) of Fig.~6 in Ref.~\cite{terasaki17a}).

This limitation of the (truncated) nonlinear higher RPA beyond the phase transition has led the authors of Ref.~\cite{terasaki17a} to implement their method on the basis of the UHF reference state, in close spirit with our approach. However, when truncated to $d=2$, which is similar to our 2p2h approximation, the method does not provide satisfactory behavior. This approach would certainly benefit from optimizing the underlying symmetry breaking reference state beyond choosing the UHF one, again in close similarity with what is presently done. Furthermore, the exact symmetry restoration included in our approach and not considered in Ref.~\cite{terasaki17a} is probably mandatory to obtain a fully satisfactory spectroscopy.

\section{Horizontal expansion (GCM)}
\label{sec:gcmresults}

\subsection{Set up}
\label{sec:setup}

Starting again from the GTCI ansatz (Eq.~\eqref{eq:fullansatz}), we now consider the other limit constituted by the adiabatic GCM method that mixes symmetry-projected vacua with different $\alpha$ deformations (at $\varphi=0$) without including any of the associated npnh excitations. A number $N_\alpha$ of vacua are mixed according to a selected set of $\{ \alpha_k\}$ values within the interval $[-\pi/4,+\pi/4]$. As mentioned previously, including states characterized by deformations $\pm|\alpha_k|$ with equal weights automatically performs the projection on good z-signature $\eta$ while generating approximate ground- and excited-state energies. This is at variance with the TCI method  discussed previously where the symmetry projection is achieved prior to the diagonalization.

One can study the optimal choices for $N_\alpha$ and $\{ \alpha_k\}$ that minimize the numerical effort while maximizing the accuracy on a set of chosen properties. In this respect, interested readers can refer to the corresponding discussion in Ref.~\cite{severyukhin06a}. Presently, we display results for one representative discretization strategy based on an odd number of equidistant points in the interval $]-\pi/4,+\pi/4[$. The odd number reflects (i) the wish to always include the point $\alpha = 0$ that is key to describing the normal phase optimally and (ii) the necessity to always include the pair of points $\pm|\alpha_k|$ in order to restore the z-signature. The $N_\alpha$ points are, thus, chosen according to the following procedure\footnote{Other choices can be more physically motivated such as picking the UHF minimum or the RVAP minimum as the first mesh point and enlarging the set by adding points distributed around it. This could be the option for realistic calculations based on a deformation parameter that does not belong to a bounded interval.}:
\begin{itemize}
\item pick $\alpha_1 = 0$,
\item pick $(N_\alpha-1)/2$ equidistant values in $]0,+\pi/4[$,
\item take the $(N_\alpha-1)/2$ opposite values in $]-\pi/4,0[$. 
\end{itemize}
The mesh is independent of the interaction stength and is the same for all values of $\chi$ on the figures shown below. 

Comparisons with the TCI results presented earlier should employ subspaces of equal dimensions. In the TCI calculations, including up to npnh configurations leads to a diagonalization in a subspace of dimension $n+1$ for $\eta = +$ or $\eta =-$. Compensating for the dimensionality associated with the symmetry restoration, and knowing that the value $\alpha=0$ is included by default but does not contribute for $\eta=-1$, GCM calculations are performed in a space of dimension $(N_\alpha-1)/2+1$ ($(N_\alpha-1)/2$) for $\eta = +$ ($\eta =-$). The dimensionalities associated with both methods are compared in Tab.~\ref{dimensiontable}. 
\begin{table}
  \setlength\extrarowheight{2pt}
  \begin{tabular}{|l|c|c||l|c|c|}
    \hline
    \hline
  TCI   & $\eta=+$ & $\eta=-$ & GCM & $\eta=+$ & $\eta=-$ \\
    \hline
$(0)\text{ph}_{\alpha}$     & 1 & 1 & $N_\alpha = 1$ & 1 & 0 \\    
    \hline
$(0+1)\text{ph}_{\alpha}$     & 2 & 2 & $N_\alpha = 3$ & 2 & 1 \\    
    \hline
$(0+1+2)\text{ph}_{\alpha}$     & 3 & 3 & $N_\alpha = 5$ & 3 & 2 \\    
    \hline
$(0+1+2+3)\text{ph}_{\alpha}$     & 4 & 4 & $N_\alpha = 7$ & 4 & 3 \\    
    \hline
    \hline
  \end{tabular}
  \caption{\label{dimensiontable} Dimensionality of TCI and GCM calculations (in a symmetry-restored basis).}
\end{table}

\subsection{Ground-state energy}
\label{sec:gsEGCM}

\begin{figure}[h!]
    \centering
    \includegraphics[scale=1.0]{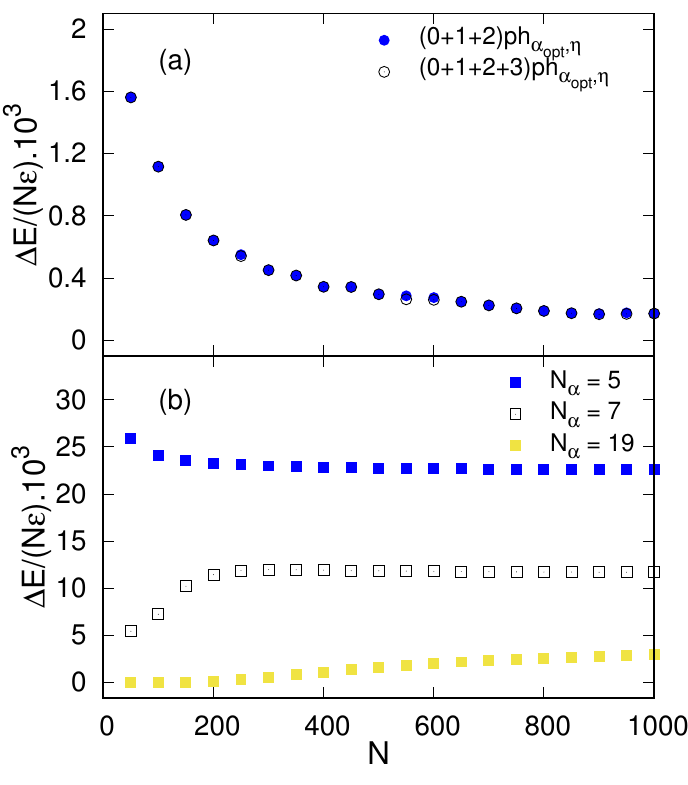}
  \caption{(Color online) Maximum error, over the interval $\chi \in [0,3]$, on the ground-state energy per particle as a function of the particle number. Panel (a): TCI method including up to  2p2h configurations (full circles) and up to  3p3h configurations (empty circles). Panel (b): horizontal GCM method based on  $N_\alpha=5,7,19$ equidistant discretization points in the interval $[-\pi/4,+\pi/4]$. }
    \label{fig:systematicsGCM}
\end{figure}

The GCM ground-state energy displayed as a function of $\chi$ can be found in Ref.~\cite{severyukhin06a} for 30 and 50 particles. Consequently, such curves are not reproduced here. We rather focus on how the maximum error on the ground-state energy per particle over the interval\footnote{This interval is obviously arbitrary but covers weak, intermediate and (rather) strong coupling regimes.} $\chi \in [0,3]$ behaves as a function of the particle number. This maximum error is displayed in Fig.~\ref{fig:systematicsGCM} for $N_\alpha=5,\ 7$ and $19$  equidistant discretization points and compared to TCI results obtained by including up to 2p2h or up to 3p3h configurations. We observe that the GCM error does not scale as the TCI one as a function of $N$ and saturates at a significantly larger value. Even though the adiabatic GCM is known to be exact in the continuous limit~\cite{ring80a}, increasing the number of equidistant mesh points reduces the error only very slowly as is illustrated for $N_\alpha=19$ in Fig.~\ref{fig:systematicsGCM}. 

\begin{figure}[h!]
    \centering
    \includegraphics[scale=1.0]{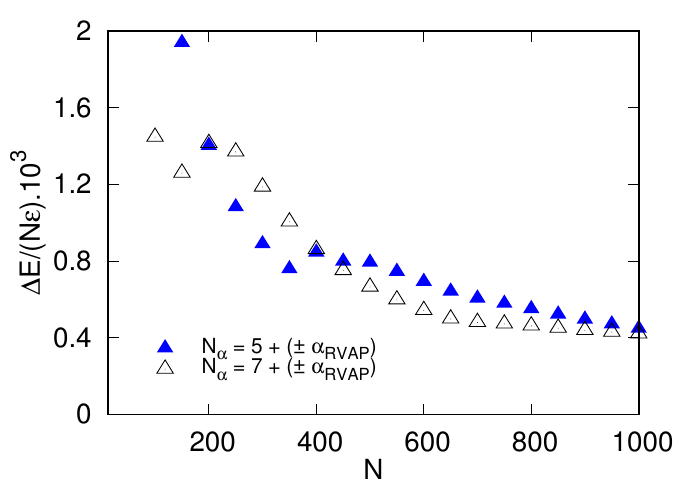}
  \caption{(Color online) Same as Panel (b) of Fig.~\ref{fig:systematicsGCM} but adding the optimal $\pm|\alpha_{\text{RVAP}}|$ mesh points to the equistant discretization corresponding to $N_\alpha=5,7$.}
    \label{fig:systematicsGCM2}
\end{figure}

To test whether this is an inherent aspect of the GCM method or a feature of our discretization strategy, the $\pm|\alpha_{\text{RVAP}}|$ mesh point is added to the equidistant discretization corresponding to $N_\alpha=5,7$. The corresponding results displayed in Fig.~\ref{fig:systematicsGCM2} convincingly demonstrate that the error is highly dependent on the discretization and that adding a single physically-optimized state does correct for the deficiency of the equidistant discretization\footnote{One notices that results are less good for $N_\alpha=7$ than for $N_\alpha=5$ for certain particle numbers. This is because the distribution of mesh points is different in both cases and may accidentally provide better results for $N_\alpha=5$ in spite of the smaller number of points. }. Eventually, the error as a function of $N$ is virtually the same as for TCI calculations.

\subsection{Excited states}
\label{sec:ciresultsesGCM}

\begin{figure}[t!]
  \centering
  \includegraphics[scale=1.0]{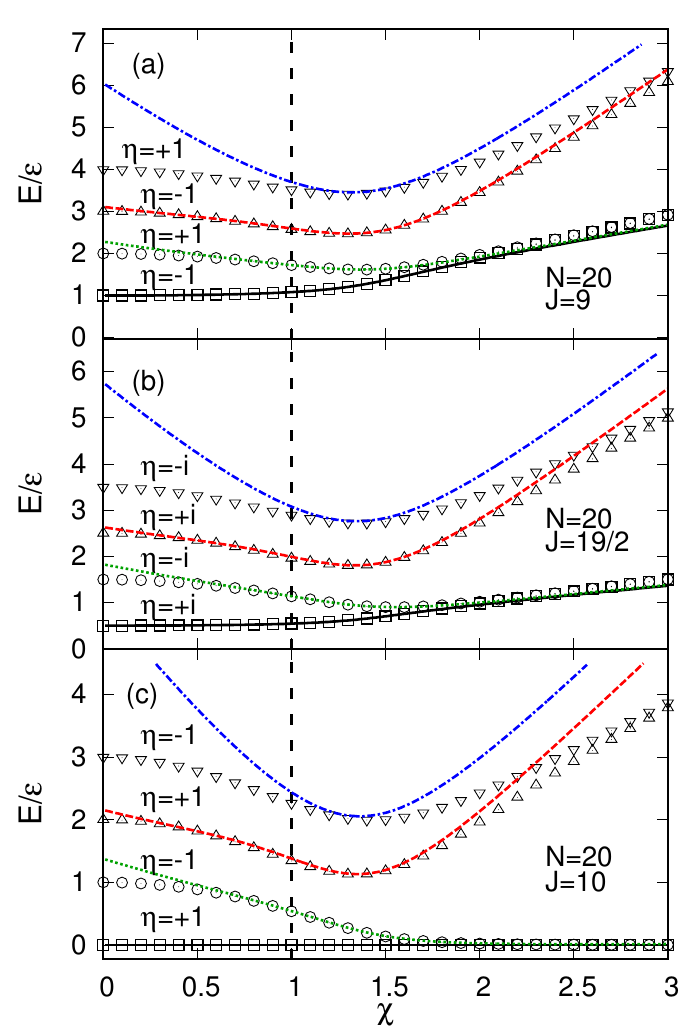}
  \caption{(Color online) Same as Fig.~\ref{fig:figure8}  with the adiabatic GCM using  $N_\alpha=5$ equidistant discretization points in the interval $[-\pi/4,+\pi/4]$.}
  \label{fig:figure9naive}
\end{figure}

The capacity of the adiabatic GCM to describe the lowest two excited states with $J=10$ has been extensively discussed in Ref.~\cite{severyukhin06a}, including a comparison to RPA results. Some of the features highlighted below\footnote{Conclusions are not strictly the same because discretization strategies are not identical in both studies.} have thus already been identified in Ref.~\cite{severyukhin06a}. The present study, in particular, extends the analysis to a larger set of excited states, among which are states with $J< 10$.

Fig.~\ref{fig:figure9naive} displays the four lowest eigenenergies for each value $J=10, 19/2, 9$ as a function of $\chi$ for $N_\alpha=5$. Comparing with Fig.~\ref{fig:figure8} , the lowest $\eta=+$ state is equally well reproduced for all couplings for $J=10$. For $J=19/2,9$, it is even slightly better described at strong coupling. The same features are seen for the lowest $\eta=-$ although it is slightly less well reproduced at small coupling because the effective dimensionality is equal to $N_\alpha-1=4$ (see Tab.~\ref{dimensiontable}) in this case. The second $\eta=-$ state is significantly less well reproduced at small coupling than in TCI calculation, in part again because of the reduced dimensionality but also because of the associated absence of active point at low deformation. The employed discretization does not capture the 3p3h nature of the second $\eta=-$ state with $J=10$ (and similarly for other $J$) in the non-interacting limit. Contrarily, the 2p2h character of the second $\eta=+$ state is well captured at low coupling thanks to the presence of the $\alpha_1=0$ mesh point. See Ref.~\cite{severyukhin06a} for a related discussion. At intermediate coupling, the second $\eta=+$ state is better reproduced than with the TCI but it drifts away at high coupling, just as the second $\eta=-$ state does.

\begin{figure}[h!]
  \centering
  \includegraphics[scale=1.0]{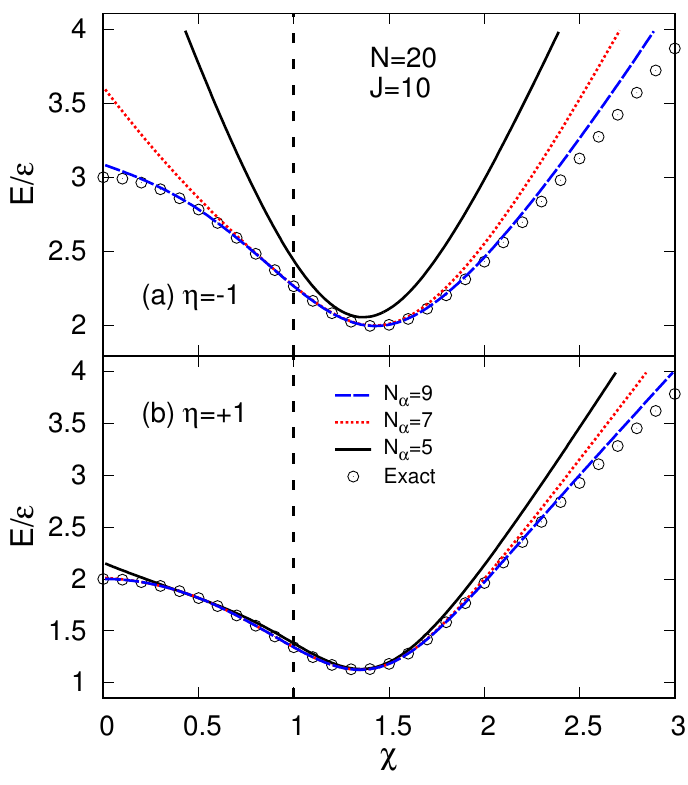}
  \caption{(Color online) Same as Fig.~\ref{fig:3p3h} for the adiabatic GCM. Results are displayed for $N_\alpha =5$ (black solid line), $7$ (red dotted line) and $9$ (blue dashed line). Exact energies are shown as open circles.} 
  \label{fig:figure9convnaive}
\end{figure}

The convergence of the GCM results as a function of the $N_\alpha$ equidistant mesh points is illustrated in Fig.~\ref{fig:figure9convnaive} for the second and third excited states with $J=10$. This can be directly compared to Fig.~\ref{fig:3p3h}.
The adiabatic GCM being exact in the continuous limit, the convergence towards exact results is no surprise. Still, energies converge rather rapidly to the exact ones and are essentially perfect over the full interval $\chi \in [0,3]$ for $N_\alpha=9$ equidistant points. By comparison with the error on the ground-state energy discussed in Sec.~\ref{sec:gsEGCM}, this demonstrates that excitation energies are much less sensitive to the discretization strategy. In comparison to the TCI method, the convergence as a function of the dimensionality is slightly better at intermediate coupling (especially for the $\eta=+$ state) but slower at small and large coupling strengths.

\section{Conclusions}
\label{sec:conclusions}

The long-term goal of the present work is to systematically access the spectroscopy of open-shell nuclei in an ab initio fashion while controlling the associated numerical cost. This is presently done by designing a novel many-body method that combines the merit of breaking and restoring symmetries with those brought about by low-rank individual excitations.

The truncated configuration interaction method based on optimized symmetry-broken and -restored states, which we denote as a purely {\it vertical} expansion, was already applied to the Richardson Hamiltonian~\cite{ripoche17a} in connection to $U(1)$ gauge symmetry. It is presently extended and applied to the Lipkin Hamiltonian in connection to the z-signature symmetry. While z-signature symmetry is close in spirit to discrete parity symmetry, this extension constitutes a step towards testing the method against the continuous rotational symmetry associated with the $SU(2)$ group.

The breaking and restoration of z-signature symmetry along with the optimization of the former \textit{in presence of} low-rank particle-hole excitations lead to highly accurate ground-state and low-lying excitation energies all the way from weak to strong coupling regimes and accross the phase transition in between. The overall quality of the approximation is very satisfactory from small to large particle numbers. In the end, it is superior in the strong coupling regime to methods like the self-consistent random-phase approximation or the self-consistent second random-phase approximation, even when the latter is implemented over a symmetry breaking vacuum. The performance of the approach is further compared to the adiabatic generator coordinate method, which we denote as a purely {\it horizontal} expansion and which is shown to provide results of similar quality.

The rationale of the present work is to eventually achieve the best possible optimization of the highly-truncated subspace employed by combining the benefits of both horizontal and vertical expansions. In the present case, however, the Lipkin Hamiltonian happened to be too simplistic to test such a highly tuned ansatz. When dealing with a richer Hamiltonian later on, the optimized combination of both expansions can be envisioned and tested.

\section*{Acknowledgement}
\label{acknow}

The authors thank A. Tichai for the proofreading of the manuscript. One of the authors (D.L.) thanks P. Schuck for useful discussions during the elaboration of this work. 

\appendix

\section{Matrix elements computation}

\subsection{Reduction of the problem}

To apply the GTCI method, one needs to compute matrix elements of an operator $O$ commuting with $R_z$, such as $H$, between $|\Psi^{JM}\rangle$ states
\begin{equation}
    \label{eq:matrixelementnotdeformed}
    \ocal^{J}_{MM'} \equiv \langle \Psi^{JM} | O | \Psi^{JM'} \rangle \, ,
\end{equation}
between symmetry breaking states
\begin{align}
    \label{eq:matrixelementdeformed}
    \ocal^{J}_{MM'}(\Omega;\Omega') &\equiv \langle \Psi^{JM} (\Omega) | O | \Psi^{JM'} (\Omega') \rangle \nonumber \\
    &= \left[ \rcal^\dagger (\Omega) \ocal \rcal (\Omega') \right]^J_{MM'} \, ,
\end{align}
and between symmetry-restored states
\begin{align}
    \label{eq:matrixelementprojected}
    {\cal O}^{J\eta}_{MM'}(\Omega;\Omega') &\equiv \langle \Theta^{JM\eta} (\Omega)| O | \Theta^{JM'\eta} (\Omega') \rangle \nonumber \\
    &= \left[ \rcal^\dagger (\Omega) \ocal \pcal_\eta \rcal (\Omega') \right]^J_{MM'},
\end{align}
where the fact that $P_\eta$ is a projector and commutes with $O$ was used. Inserting the expression of the z-signature projector (Eq.~\eqref{eq:projector}) in the symmetry-restored matrix elements (Eq.~\eqref{eq:matrixelementprojected}) actually allows one to express them in terms of the non-projected ones (Eq.~\eqref{eq:matrixelementdeformed}) according to
\begin{align}
    \ocal^{J\eta}_{MM'}(\Omega;\Omega') = &\frac{1}{2} \left[ \ocal^{J}_{MM'}(\Omega;\Omega') \right.\nonumber \\
    & \left. \,\, + \eta \eta_{M'} \ocal^{J}_{MM'}(\Omega;-\Omega')\right] \, , \label{rewrittenMA}
\end{align}
where $\eta_{M'}$ is the z-signature of $| \Psi^{JM'} \rangle$.
Furthermore, defining the rotated operator $O(\Omega)$ as
\begin{equation}
    \label{eq:rotoperator}
    O(\Omega) \equiv R^\dagger (\Omega) O R (\Omega),
\end{equation}
one can rewrite the matrix elements between symmetry-breaking states, which are the key inputs to Eq.~\ref{rewrittenMA}, as
\begin{align}
    \label{eq:matrixelementdeformed2}
    \ocal^{J}_{MM'}(\Omega;\Omega') &= \left[ \ocal (\Omega) \rcal^\dagger (\Omega) \rcal (\Omega') \right]^{J}_{MM'} \\
    &= \sum_{M''} \ocal (\Omega)^{J}_{MM''} \left[ \rcal^\dagger (\Omega) \rcal (\Omega') \right]^{J}_{M''M'}, \nonumber
\end{align}
where a completeness relation was inserted.  Eventually, the problem reduces to the computation of the matrix elements of the rotated operator $O(\Omega)$ and of the rotation operator $ R^\dagger (\Omega)  R (\Omega')$ between $|\Psi^{JM}\rangle$ states. The matrix elements of the rotation operator are Wigner D-functions $D^{J}_{MM'}(\alpha,\beta,\gamma)$, where identities relating variables $(\Omega,\Omega')$ to variables $(\alpha,\beta,\gamma)$ exist~\cite{varshalovich88a}. The summation over $M''$ in Eq.~\eqref{eq:matrixelementdeformed2} runs in principle from $-J$ to $+J$ but properties of $O$ can greatly reduced this range, e.g. $M''=M$ for $O=1$ and $|M''-M| \leq 2$ for $O=H$. This is easily checked from the expressions of the rotated norm overlap
\begin{equation}
    \ncal (\Omega)^J_{MM'} = \delta_{MM'} \, ,
\end{equation}
and of the rotated Hamiltonian matrix elements \eqref{eq:rotatedhamiltonianme} given in the next section.

\subsection{Matrix elements of $H(\Omega)$}

The rotated Hamiltonian is expressed as a function of rotated quasi-spin operators according to
\begin{align}
H(\Omega) &= R^\dagger(\Omega) H R(\Omega) \nonumber \\
		  &= \epsilon J_z (\Omega) - \frac{V}{2} \left[ J^2_+ (\Omega) + J^2_- (\Omega) \right].
\end{align}
The rotated operator $J_z (\Omega)$ is expressed as a function of unrotated quasi-spin operators using Eq.~\eqref{eq:j0pmexpression} and Eq.~\eqref{eq:quasispinrotation}
\begin{align}
J_z (\Omega) &= R^\dagger(\Omega) J_z R(\Omega) \nonumber \\ 
    &= \frac{1}{2} \sum_{p\sigma} \sigma \, a^\dagger_{p\sigma}(\Omega) a_{p\sigma}(\Omega) \\[-8pt]
    &= \cos(2 \alpha) J_z - \frac{1}{2} \sin(2 \alpha) \left[ e^{+i\varphi}J_+ + e^{-i\varphi} J_-  \right]. \nonumber
\end{align}
In a similar manner one obtains
\begin{align}
    &J^2_+ (\Omega) + J^2_- (\Omega) = \sin^2(2 \alpha) \cos(2\varphi) \left( 3 J^2_z - J^2 \right) \nonumber \\
    &~~~~~~+ \left( \cos^4 (\alpha) + \sin^4 (\alpha) e^{+4i\varphi} \right) J^2_+ \nonumber \\
    &~~~~~~+  h.c. \\
    &~~~~~~+ \left( \cos^2 (\alpha) - \sin^2 (\alpha) e^{+4i\varphi} \right) \sin(2 \alpha) e^{-i\varphi} \{J_z,J_+\} \nonumber \\
    &~~~~~~  + h.c. \nonumber
\end{align}
where $\{J_z,J_\pm\} = J_z J_\pm + J_\pm J_z$. Knowing the action of unrotated quasi-spin operators on $|\Psi^{JM}\rangle$
\begin{subequations}
\begin{align}
    J^2 |\Psi^{JM}\rangle &= J(J+1) |\Psi^{JM}\rangle \, , \\
    J_\pm |\Psi^{JM}\rangle &= \sqrt{J(J+1)-M(M\pm 1)} |\Psi^{JM\pm 1}\rangle \, , \\
    J_z |\Psi^{JM}\rangle &= M |\Psi^{JM}\rangle \, ,
\end{align}
\end{subequations}
one obtains matrix elements of the rotated Hamiltonian in the $|\Psi^{JM}\rangle$ basis under the form
\begin{widetext}
\begin{align}
\label{eq:rotatedhamiltonianme}
    \hcal (\Omega)^J_{MM'} &= +~\epsilon \cos(2 \alpha) M~\delta_{MM'} -\frac{V}{2} \sin^2(2 \alpha) \cos(2\varphi) \left( 3 M^2 - J(J + 1) \right) \delta_{M M'} \nonumber \\
    &~~~-\frac{\epsilon}{2}~\sum_{\sigma = \pm 1} \sin(2 \alpha) e^{i\sigma\varphi} \sqrt{J(J+1) - M(M-\sigma)}~\delta_{M M' + \sigma} \nonumber \\[-6pt]
    &~~~-\frac{V}{2} \sum_{\sigma = \pm 1} \left( \cos^2 (\alpha) - \sin^2 (\alpha) e^{4i\sigma\varphi} \right) \sin(2 \alpha) e^{-i\sigma\varphi} (2M - \sigma) \sqrt{J(J+1) - M(M-\sigma)}~\delta_{M M' + \sigma} \nonumber \\[-6pt]
    &~~~-\frac{V}{2} \sum_{\sigma = \pm 1} \left( \cos^4 (\alpha) + \sin^4 (\alpha) e^{4i\sigma\varphi} \right) \sqrt{(J(J+1) - (M-\sigma)^2)^2 - (M-\sigma)^2}~\delta_{M M' + 2\sigma} \, .
\end{align}
\end{widetext}


\begin{thebibliography} {99}
  \bibitem{navratil00a} P. Navr\'atil, J. P. Vary and B. R. Barrett, Phys. Rev. Lett. {\bf 84}, 5728 (2000)
  \bibitem{pieper02a} S. C. Pieper, K. Varga and R. B. Wiringa, Phys. Rev. C {\bf 66}, 044310 (2002)
  \bibitem{kowalski04a} K. Kowalski, D. J. Dean, M. Hjorth-Jensen,  T. Papenbrock and P. Piecuch, Phys. Rev. Lett. {\bf 92}, 132501 (2004)
\bibitem{hagen14a} G. Hagen, T. Papenbrock, M. Hjorth-Jensen, D.J. Dean, Rep. Prog. Phys. {\bf 77}, 096302 (2014)
  \bibitem{roth07a} R. Roth and P. Navr\'atil, Phys. Rev. Lett. {\bf 99}, 092501 (2007)
  \bibitem{epelbaum10a} E. Epelbaum, H. Krebs, D. Lee and U.-G. Mei{\ss}ner, Phys. Rev. Lett. {\bf 104}, 142501 (2010)
 
\bibitem{dickhoff04a} W. H. Dickhoff, C. Barbieri, Prog. Part. Nucl. Phys. {\bf 52}, 377 (2004)

  \bibitem{cipollone13a} A. Cipollone, C. Barbieri and P. Navr\'atil, Phys. Rev. Lett. {\bf 111}, 062501 (2013)
  \bibitem{hergert16a} H. Hergert, S. K. Bogner, T. D. Morris, A. Schwenk and K. Tsukiyama, Phys. Rep. {\bf 621}, 165 (2016)
\bibitem{hergert17a} H. Hergert, Phys. Scr. {\bf 92}, 023002 (2017)
\bibitem{launey16a} K. D. Launey, T. Dytrych, J. Draayer, Prog. Part. Nucl. Phys. {\bf 89}, 101 (2016)
\bibitem{gebrerufael16a} E. Gebrerufael, K. Vobig, H. Herbert, R. Roth, Phys. Rev. Lett. {\bf 118}, 152503 (2017)    
\bibitem{tichai17a} A. Tichai, E. Gebrerufael, R. Roth, arXiv:1703.05664  

  \bibitem{soma11a} V. Som\`a, T. Duguet and C. Barbieri, Phys. Rev. C {\bf 84}, 064317 (2011)
  \bibitem{soma14b} V. Som\`a, A. Cipollone, C. Barbieri, P. Navr\'atil and T. Duguet, Phys. Rev. C {\bf 89}, 061301(R) (2014)
  \bibitem{henderson14a} T. M. Henderson, G. E. Scuseria, J. Dukelsky, A. Signoracci and T. Duguet, Phys. Rev. C {\bf 89}, 054305 (2014)
  \bibitem{duguet15a} T. Duguet, J. Phys. G: Nucl. Part. Phys. {\bf 42}, 025107 (2015)
  \bibitem{signoracci15a} A. Signoracci, T. Duguet, G. Hagen and G. R. Jansen, Phys. Rev. C {\bf 91}, 064320 (2015)
  \bibitem{duguet17a} T. Duguet and A. Signoracci, J. Phys. G: Nucl. Part. Phys. {\bf 44}, 015103 (2017)

\bibitem{bogner14a} S. K. Bogner, H. Hergert, J. D. Holt, A. Schwenk, S. Binder, A. Calci, J. Langhammer, R. Roth, Phys. Rev. Lett. {\bf 113}, 142501 (2014) 
\bibitem{jansen14a} G. R. Jansen, J. Engel, G. Hagen, P. Navratil, A. Signoracci, Phys. Rev. Lett. {\bf 113}, 142502 (2014)
\bibitem{dikmen15a} E. Dikmen, A. F. Lisetski, B. R. Barrett, P. Maris, A. M. Shirokov, J. P. Vary, Phys. Rev. C {\bf 91}, 064301 (2015) 

  \bibitem{schmid84a} K. W. Schmid, F. Gr\"ummer and A. Faessler, Nucl. Phys. {\bf A431}, 205 (1984)

  \bibitem{hara95a} K. Hara and Y. Sun, Int. Jour. of Mod. Phys. E {\bf 4}, 637-785 (World Scientific, 1995)

  \bibitem{allaart88a} K. Allaart, E. Boeker, G. Bonsignori, M. Savoia and Y. Gambhir, Phys. Rep. {\bf 169}, 209 (1988)
  \bibitem{caprio12a} M. A. Caprio, F. Q. Luo, K. Cai, C. Constantinou and V. Hellemans, J. Phys. {\bf G39}, 105108 (2012)

  \bibitem{lacroix12a} D. Lacroix and D. Gambacurta, Phys. Rev. C {\bf 86}, 014306 (2012)
  \bibitem{gambacurta12a} D. Gambacurta and D. Lacroix, Phys. Rev. C {\bf 86}, 064320 (2012)
  \bibitem{ripoche17a} J. Ripoche, D. Lacroix, D. Gambacurta, J.-P. Ebran and T. Duguet, Phys. Rev. C {\bf 95}, 014326 (2017) 


  \bibitem{baerdemacker17a} S. De Baerdemacker, P. W. Claeys, J.-S. Caux, D. Van Neck and P. W. Ayers, arxiv:1712.01673

  \bibitem{lipkin65a} H. J. Lipkin, N. Meshkov and A. J. Glick, Nucl. Phys. A {\bf 62}, 188 (1965)
  \bibitem{meshkov65a} N. Meshkov, A. J. Glick and H. J. Lipkin, Nucl. Phys. A {\bf 62}, 199 (1965)
  \bibitem{glick65a} A. J. Glick, H. J. Lipkin and N. Meshkov, Nucl. Phys. A {\bf 62}, 211 (1965)
  \bibitem{agassi66a} D. Agassi, H. J. Lipkin and N. Meshkov, Nucl. Phys. A {\bf 86}, 321 (1966)
  \bibitem{schuck73a} P. Schuck and S. Ethofer, Nucl. Phys. A {\bf 212}, 269 (1973)

  
  \bibitem{holzwarth73a} G. Holzwarth, Nucl. Phys. A {\bf 207}, 545 (1973)
  \bibitem{ring80a} P. Ring and P. Schuck, The Nuclear Many-Body Problem (Springer-Verlag, New-York, 1980)
  \bibitem{zhang90a} W.-M. Zhang, D. H. Feng and R. Gilmore, Rev. Mod. Phys. {\bf 62}, 867 (1990)
  \bibitem{dukelsky90a} J. Dukelsky and P. Schuck, Nucl. Phys. A {\bf 512}, 466 (1990)
  \bibitem{dukelsky91a} J. Dukelsky and P. Schuck, Mod. Phys. Lett. A {\bf 6}, 2429 (1991)
  \bibitem{robledo92a} L. M. Robledo, Phys. Rev. C {\bf 46}, 238 (1992)
  \bibitem{dukelsky99a} J. Dukelsky, J. G. Hirsch and P. Schuck, Eur. Phys. J. A {\bf 7}, 155 (2000)
  \bibitem{delion05a} D. S. Delion, P. Schuck and J. Dukelsky, Phys. Rev. C {\bf 72}, 064305 (2005)
  \bibitem{severyukhin06a} A. P. Severyukhin, M. Bender, and P.-H. Heenen, Phys. Rev. C {\bf 74}, 024311 (2006)
  \bibitem{hermes17a} M. R. Hermes, J. Dukelsky and G. E. Scuseria, Phys. Rev. C {\bf 95}, 064306 (2017)
  \bibitem{wahlen17a} J. M. Wahlen-Strothman, T. M. Henderson, M. R. Hermes, M. Degroote, Y. Qiu, J. Zhao, J. Dukelsky and G. E. Scuseria, J. Chem. Phys. {\bf 146}, 054110 (2017)

  \bibitem{rodriguez05a} T. R. Rodriguez, J. L. Egido and L. M. Robledo, Phys. Rev. C {\bf 72}, 064303 (2005)

  \bibitem{terasaki17a} J. Terasaki, A. Smetana, F. Simkovic and M. I. Krivoruchenko, arXiv:1701.08368
  \bibitem{schuck16a}  P. Schuck and M. Tohyama, Phys. Rev. B {\bf 93}, 165117 (2016).
  
  \bibitem{varshalovich88a} D. A. Varshalovich, A. N. Moskalev and V. K. Khersonskii, Quantum theory of angular momentum (World Scientific, Singapore, 1988)

\end{thebibliography}
\end{document}